\documentclass[conference]{IEEEtran}
\usepackage{cite}
\usepackage{amsmath,amssymb,amsfonts}
\usepackage{algorithmic}
\usepackage{graphicx}
\usepackage{textcomp}
\usepackage{xcolor}
\usepackage{paralist}
\usepackage{subcaption}
\usepackage{xspace}

\usepackage{listings}
\usepackage{float}
\newfloat{listing}{tbhp}{lst}
\floatname{listing}{Listing}

\definecolor{mGreen}{rgb}{0,0.6,0}
\definecolor{mGray}{rgb}{0.5,0.5,0.5}
\definecolor{mPurple}{rgb}{0.58,0,0.82}
\definecolor{backgroundColour}{rgb}{0.95,0.95,0.92}
\newcommand{\system}{\textsc{Heats}\xspace}
\newcommand{\sys}{\system}

\lstdefinestyle{CStyle}{
	basicstyle=\tiny,
	escapeinside={<@}{@>},
    backgroundcolor=\color{backgroundColour},   
    commentstyle=\color{mGreen},
    keywordstyle=\color{magenta},
    numberstyle=\tiny\color{mGray},
    stringstyle=\color{mPurple},
    breakatwhitespace=false,         
    breaklines=true,                 
    captionpos=b,                    
    keepspaces=true,                 
    numbers=left,                    
    numbersep=5pt,                  
    showspaces=false,                
    showstringspaces=false,
    showtabs=false,                  
    tabsize=2,
    language=C
}

\def\BibTeX{{\rm B\kern-.05em{\sc i\kern-.025em b}\kern-.08em
    T\kern-.1667em\lower.7ex\hbox{E}\kern-.125emX}}
\begin{document}

\title{LEGaTO: Low-Energy, Secure, and Resilient Toolset for Heterogeneous Computing
}

\author{\IEEEauthorblockA{ 
B. Salami,
K. Parasyris,
A. Cristal, 
O. Unsal, 
X. Martorell, 
P. Carpenter, 
R. De La Cruz, 
L. Bautista,
D. Jimenez, \\ 
C. Alvarez, 
S. Nabavi,  
S. Madonar (BSC),
M. Pericàs,
P. Trancoso,
M. Abduljabbar,
J. Chen,
P. N. Soomro,\\
M Manivannan (Chalmers),
M. Berge,
S. Krupop (CHR),
F. Klawonn, Al Mekhlafi, S. May (HZI),\\
T. Becker, G. Gaydadjiev (Maxeler),
D. Odman, H. Salomonsson, D. Dubhashi (MIS) ,
O. Port, Y.  Etsion (Technion),\\
Le Quoc Do, Christof Fetzer (TUD), 
M. Kaiser, N. Kucza, J. Hagemeyer, R. Griessl, L. Tigges, K. Mika,\\ A. Hüffmeier, Th. Jungeblut (UBI),
M. Pasin, V. Schiavoni, I. Rocha, C. Göttel and P. Felber (UniNE)
}}
 
\maketitle

\begin{abstract}
The LEGaTO project leverages task-based programming models to provide a software ecosystem for Made in-Europe heterogeneous hardware composed of CPUs, GPUs, FPGAs and dataflow engines. The aim is to attain one order of magnitude energy savings from the edge to the converged cloud/HPC, balanced with the security and resilience challenges. LEGaTO is an ongoing three-year EU H2020 project started in December 2017.
\end{abstract}

\section{Introduction}
\label{sec:introduction}
In the last couple of decades, technological advances in the ICT sector have been the dominant factors in global economic growth, not to mention an increase in the quality of life for billions of people. At the heart of this advance lies Moore’s Law, which states that the number of transistors in an integrated chip will double every 18 to 24 months with each step in the silicon manufacturing technology node. However, due to the fundamental limitations of scaling at the atomic scale, coupled with heat density problems of packing an ever-increasing number of transistors in a unit area, Moore’s Law has slowed down in the last two years or so and will soon stop altogether \cite{ITRS}. The implication is that, in the future, the number of transistors that could be incorporated into a processor chip will not increase. This development threatens the future of the ICT sector as a whole. 
As  a  solution  to  this  challenge, there has recently been a dramatic increase in efforts toward heterogeneous computing, including the integration of heterogeneous cores on die, utilizing general-purpose GPUs and combining CPUs, GPUs and FPGAs in integrated SoCs.

Heterogeneity aims to solve the problems associated with the end of Moore’s Law by incorporating more specialized compute units in the system hardware and by utilizing the most efficient compute unit for each computation. However, while software-stack support for heterogeneity is relatively well developed for performance, it is severely lacking for power- and energy-efficient computing. Given that the ICT sector is responsible for ~5\% of global electricity consumption \cite{van2014trends}, 
software stack support for energy-efficient heterogeneous computing is critical to the future growth of the ICT industry. The primary ambition of the LEGaTO project is to address this challenge by starting with a Made-in-Europe mature software stack and by optimizing this stack to support energy-efficient computing on a commercial cutting-edge European developed CPU–GPU–FPGA heterogeneous hardware substrate \cite{griessl2014scalable} and FPGA-based Dataflow Engines (DFE), which will lead to an order of magnitude increase in energy efficiency. The LEGaTO project will utilize a completely integrated software system stack supporting generalized tasks for low-energy, secure and reliable parallel computing. We foresee that optimization opportunities for low-energy computing can be maximized through the task abstraction. 

\begin{figure}
\centering
\includegraphics[width=0.5\textwidth]{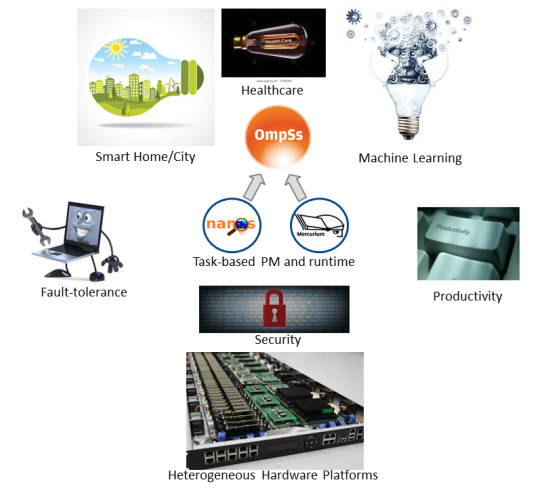}
\caption{LEGaTO ecosystem.}
\label{fig:legato-eco}
\vspace{-1.5em}
\end{figure}

Also, it is important to balance the advantage of a low-energy heterogeneous CPU/FPGA/GPU hardware platform with security and resilience challenges. We are therefore working on ensuring the resilience of the software stack running on this hardware, while simultaneously optimizing for performance and low power. For fault tolerance we would like to exploit the unique characteristics of the heterogeneous CPU/GPU/FPGA platform in the runtime; for example by replicating tasks intelligently on diverse processing elements exploiting the spatial/temporal slack; additionally, we will investigate energy-efficient selective replication where only the most reliability-critical tasks will be replicated. Furthermore, we will leverage the task programming model for detecting error propagation across task boundaries and walking the task dependency graph at runtime, which will help with failure root cause analysis. Finally, we will use the properties of the task model to design application-level energy-efficient checkpointing where only the necessary and sufficient data (declared at the task entry) will be checkpointed. For security, we will develop energy-efficient security-by-design by leveraging instruction-level hardware support for security (SGX in x86 and TrustZone in ARM) to accelerate software-based security implementations. The LEGaTO ecosystem is shown in Fig. \ref{fig:legato-eco}.

In the rest of the paper, we will briefly introduce the LEGaTO full stack ecosystem including hardware platforms, runtime and middleware system, compiler and programming models, and use cases (Section \ref{sec:overall}). Then for each LEGaTO abstraction layers, we will highlight a representative technology already-developed during the project. More specifically, we elaborate on the hardware-level aggressive undervolting technology for FPGAs (Section \ref{sec:undervolting}), middleware-level GPU checkpointing (Section \ref{sec:checkpointing}), compiler-level task-oriented and energy-aware orchestrator for heterogeneous clusters or HEATS (Section \ref{sec:heats}), and finally Smart Mirror as one of the LEGaTO use cases (Section \ref{sec:smartmirror}). In Section \ref{sec:future}, we will briefly mention the ongoing efforts to the end of the project. 

\begin{figure}
\includegraphics[width=0.5\textwidth]{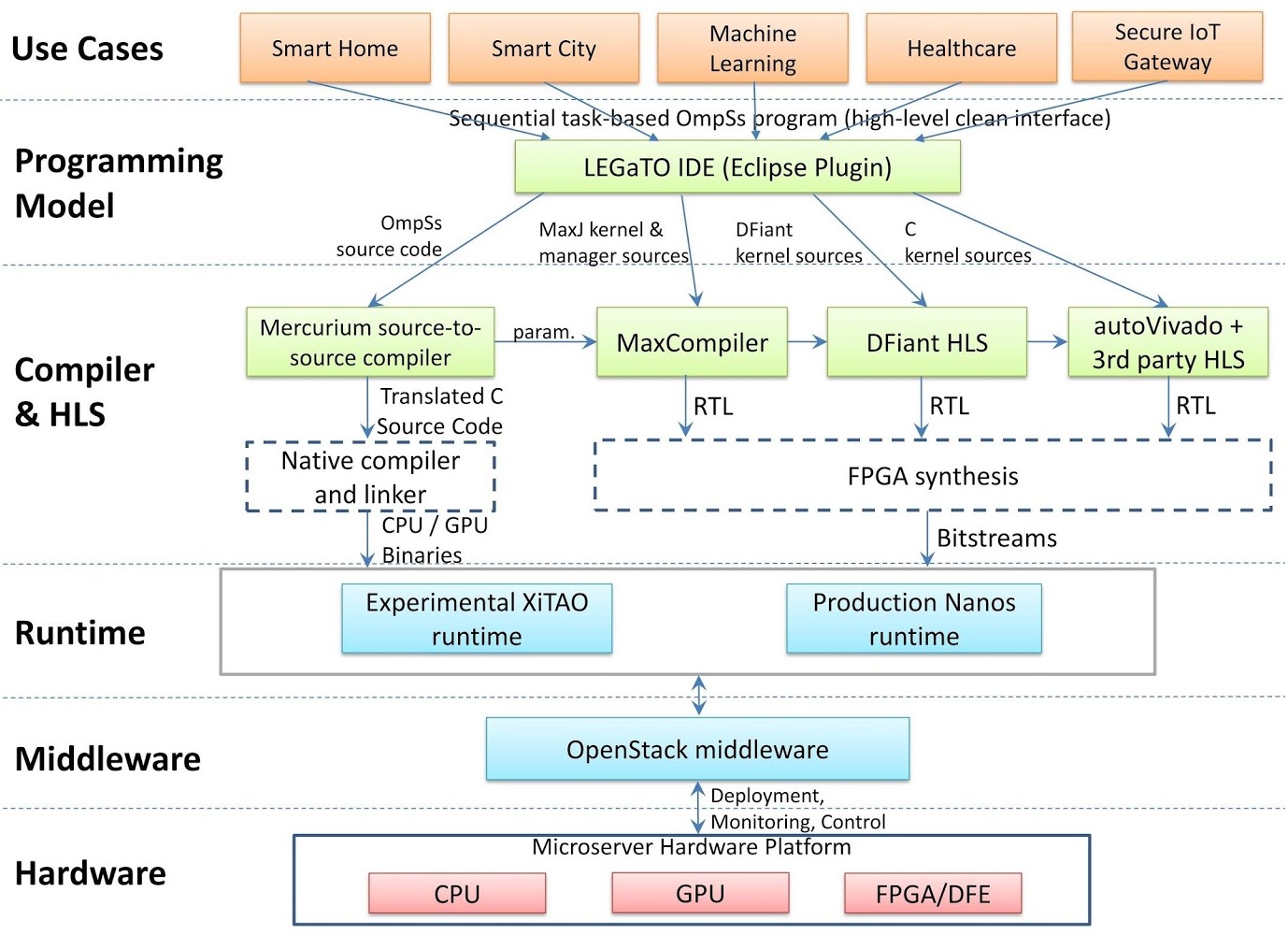}
\caption{LEGaTO hardware-software stack.} 
\label{fig:legato-stack}
\vspace{-1.5em}
\end{figure} 

\section{The LEGaTO Approach}
\label{sec:overall}
In Fig. \ref{fig:legato-stack}, we present the LEGaTO-ecosystem, on the higher level, namely the application use cases, LEGaTO targets a wide variety of application domains such as machine learning, Smart Homes and healthcare. These applications have a different set of requirements in terms of energy efficiency, Fault Tolerance, and Security. All these requirements will be facilitated by a single programming model which increases the productivity of the development process and allows the developer to specify their requirements. At execution time two runtime systems will be responsible to satisfy the user requirements. The runtime systems will reduce the energy efficiency of the application by scheduling the computations to the most energy-efficient device of the heterogeneous hardware architecture. We provide more details below.

\subsection{Hardware}
\begin{figure}
\begin{center}
\includegraphics[width=0.4\textwidth]{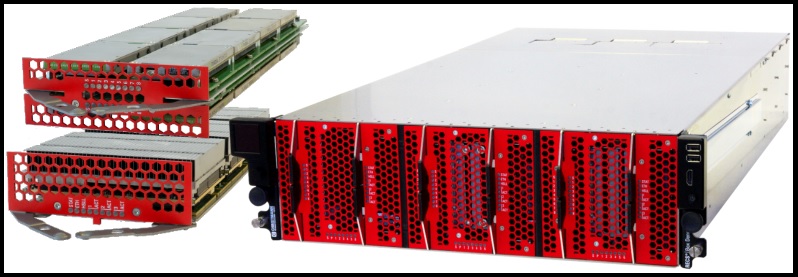}
\caption{LEGaTO hardware platform RECS$|$BOX: Heterogeneous microserver platform with carriers for low-power and high-performance microserver modules} 
\label{fig:RECS-box}
\end{center}
\vspace{-1.5em}
\end{figure}

\begin{figure}[b]
\begin{center}
\includegraphics[width=0.5\textwidth]{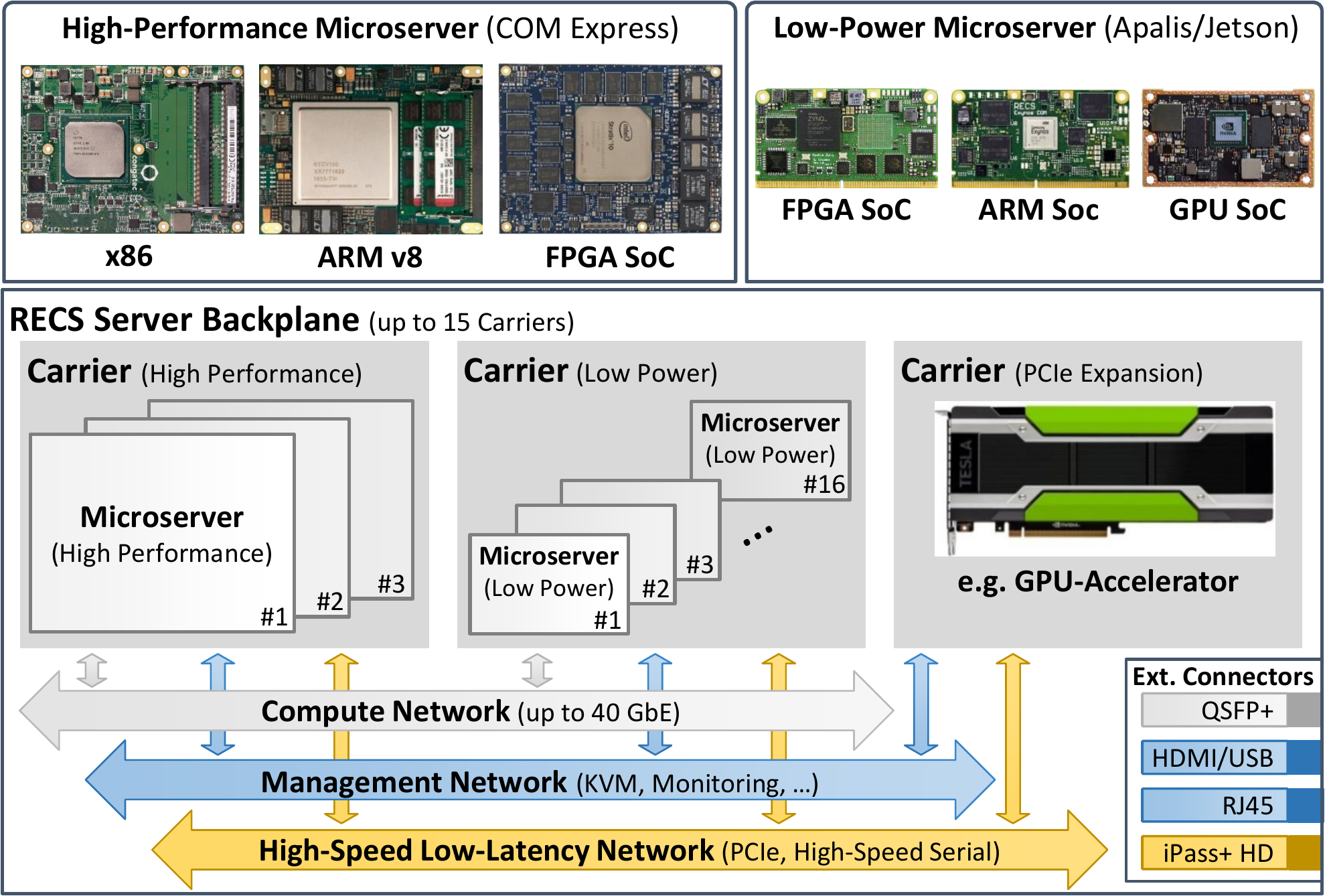}
\caption{High-level architecture of the LEGaTO hardware platform} 
\label{fig:RECS-arch}
\end{center}
\end{figure}

The RECS$|$BOX platform used in LEGaTO supports the full range of heterogeneous microserver technology in high-performance as well as low-power variants (see Fig.~\ref{fig:RECS-box}). The server supports up to 144 heterogeneous, modular microserver nodes like CPU (x86 or ARM64), GPU, FPGA and SoCs in a compact 3~RU form factor. Due to the modular approach the hardware platform provides an optimal match for a wide range of use cases and offers the ability to tightly interconnect the microservers via a flexible high-speed, low-latency communication infrastructure (see Fig. \ref{fig:RECS-arch}). More details can be found in~\cite{oleksiak2017m2dc}. 
In addition to the cloud platform, a scalable, heterogeneous edge platform is developed within the project, supporting microservers for cloud as well as edge use cases. The modular approach allows to tailor the platform towards the specific use cases, an example is provided in section \ref{sec:smartmirror}.


\subsection{Middleware}
The management and composition of resources is one main task of the LEGaTO middleware layer. This requires a good understanding of the available hardware and its configuration. A suitable middleware layer is required to abstract out the complexity of the various hardware management possibilities. This will also improve the user-level experience. Toward this goal, in LEGaTO an embedded firmware is running on management CPUs within the hardware, managing, controlling and monitoring it on a low-level. The other main block of the LEGaTO middleware is OpenStack, which is an open source software platform for managing cloud computing with the idea of providing infrastructure as a service.

\subsection{Runtime}
The main task of the runtime is to make efficient use of the underlying hardware by smart scheduling of tasks across different resources. LEGaTO is based on two runtime systems, including OmpSs \cite{duran2011ompss} and XiTAO \cite{pericas2016poster}. OmpSs is based on task parallelism, and very similar to OpenMP tasking. It is being used as a forerunner prototyping environment for future OpenMP features. On GPUs, both CUDA and OpenCL kernels are supported. For FPGAs, OmpSs uses the vendor IP generation tools (Xilinx Vivado and Vivado HLS, or Altera Quartus), to generate the hardware configuration from high-level code. Also, XiTAO is a task-based runtime that generalizes the concept of a task into a parallel computation with arbitrary (elastic) resources. By matching task requirements with hardware resources (cores, memory, etc) at runtime, XiTAO targets high parallelism and provides constructive sharing and interference freedom. Overall, this strategy improves the energy efficiency of the computation.

\subsection{Compiler and High-Level Synthesis (HLS)}
To develop applications for different hardware resources targeting objectives like energy efficiency, performance, security, reliability, or productivity, we are building a toolchain to map applications written in a high-level task-based dataflow language onto such heterogeneous platforms. 

\subsection{Programming Model}
In LEGaTO, we developed a front-end system to support applications  at compilation and runtime. In the heart of this toochain there is OmpSs programming model that allows expressing parallelism for available resources among the host SMP cores, integrated/discrete GPUs, and/or FPGAs.

\subsection{Use Cases}
In LEGaTO, we develop and optimize different real use cases with the help of the LEGaTO workflow, including Smart Home, Smart City, Infection Research, Machine Learning, and Secure IoT Gateway. These benchmarks have been already able to be optimized using one of the toolflows that the LEGaTO project provides.

\section{Hardware: FPGA Undervolting}
\label{sec:undervolting}
Aggressive undervolting, \textit{i.e.,} supply voltage underscaling below the nominal level is one of the most efficient techniques to reduce the power consumption of the chip, because dynamic power is quadratic in voltage. In addition, usually vendors add a very large voltage guardband below the nominal voltage level to guarantee the correct functionality in the worst case process and environmental conditions. This guardband is not necessary for many real-world applications and thus, eliminating it can deliver significant power saving. However, by further undervolting below the guardband level, the reliability of the underlying hardware can be affected due to the aggressive circuit delay increase. In LEGaTO, we aim to leverage the aggressive undervolting technique and initially we evaluated it for FPGAs, as described below in more detail.

\subsection{Experimental Methodology}
Experiments are performed on representative FPGAs from Xilinx, a main vendor, including VC707 (performance-oriented Virtex), two identical samples of KC705 (A \& B, power-oriented Kintex), and ZC702 (CPU-based Zynq). Among various FPGA components, a major part of experiments is initially performed on on-chip memories or Block RAMs (BRAMs), thanks to their importance in the architecture of state-of-the-art applications like FPGA-based DNNs as well as the capability of their voltage rail to be independently regulated. BRAMs are a set of small blocks of SRAMs, distributed over the chip, and in a programmable-fashion can be chained to build larger memories. All evaluated platforms are fabricated with 28nm technology and their nominal/default BRAM’s voltage level ($V_{CCBRAM}$) is 1V. 

\begin{figure}
\includegraphics[width=0.5\textwidth]{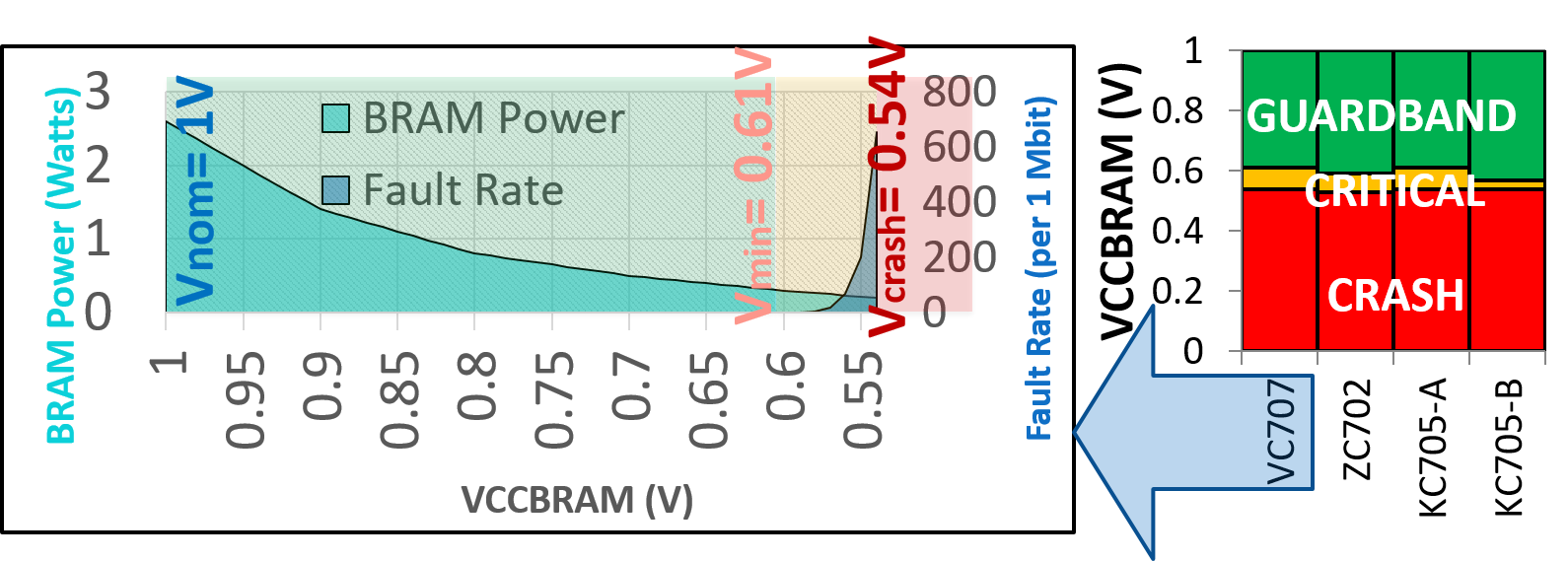}
\caption{Voltage behavior and power/reliability trade-off behavior of
FPGAs (shown for VC707 at ambient temperature).} 
\label{fig:legato-fpga}
\end{figure}

\subsection{Experimental Results}
As shown in Fig. \ref{fig:legato-fpga}, undervolting $V_{CCBRAM}$ below the nominal level, the performance or reliability of the BRAMs are not affected until a certain level, \textit{i.e.,} minimum safe voltage or $V_{min}$. This region is the guardband, which is mainly considered by vendors to ensure the worst-case environmental and process scenarios. In the guardband voltage region, data can be safely retrieved without compromising reliability.  Further undervolting, although the FPGA is still accessible, the content of some BRAMs experience faults or bit-flips. We call it as the critical region. Finally, further undervolting, the DONE pin is unset at $V_{crash}$ and the FPGA does not respond for any request in the crash region. As seen, there is a slight difference of mentioned voltage margins among platforms even for identical samples of KC705; however, those three voltage regions are recognizable for all. As shown in Fig. \ref{fig:legato-fpga} (for VC707), the power is continuously reduced through undervolting in both guardband and critical voltage regions, led to more than 90\% of power saving at $V_{crash}$ vs. $V_{nom}$. However, within the critical region, some of the memories are infected. The fault rate exponentially increases by further undervolting within the critical region up to to 652 faults/Mbit at $V_{crash}$. In the same line, we observe that the fault rate exponentially increases up to 153, 254, and 60 faults/Mbit at $V_{crash}$ for ZC702, KC705-A, and KC705-B, respectively. 

\subsection{Ongoing Work}
Our initial experimental study on the FPGAs reveals that aggressive undervolting is potentially a promising technique as in detail described in \cite{salami2018comprehensive}; thus, we aim to exploit it to achieve the major goal of the LEGaTO project, \textit{i.e.,} energy-efficiency, as well as improving the resilience of the underlying hardware as another goal of LEGaTO. Hence, we are working on the integration of the aggressive undervolting with LEGaTO software stack such as task-based low-voltage OmpSs@FPGA as well as further evaluations with the LEGaTO use cases like ML-based application. Note that due to inherent resilience of ML models \cite{salami2018resilience}, aggressive undervolting can  lead to significant power saving even below the voltage guardband region.  

\section{Middleware: GPU Checkpointing}
\label{sec:checkpointing}
\begin{listing}[t]
\begin{lstlisting}[style=CStyle]
int main(int argc, char *argv[]){
	int rank, nbProcs;
	double *h,*g;
	int i;
	MPI_Init(&argc, &argv);
	<@\textcolor{red}{FTI\_Init}@>(argv[1], MPI_COMM_WORLD);
	MPI_Comm_size(<@\textcolor{red}{FTI\_COMM\_WORLD}@>, &nbProcs);
	MPI_Comm_rank(<@\textcolor{red}{FTI\_COMM\_WORLD}@>, &rank);
	cudaMallocManaged(&h,sizeof(double)*nElements, flags);
	cudaMalloc(&g,sizeof(double)*nElements);
	initData(&h,&g);
	<@\textcolor{red}{FTI\_Protect}@>(0, &i, 1, FTI_INTG);
	<@\textcolor{red}{FTI\_Protect}@>(1, h, nElements, FTI_DBLE);
	<@\textcolor{red}{FTI\_Protect}@>(2, g, nElements, FTI_DBLE);
	for(i = 0; i < N; i++){
		<@\textcolor{red}{FTI\_Snapshot}@>();
		performComputations(h,g,i);
	}
	<@\textcolor{red}{FTI\_Finalize}@>();
	MPI_Finalize();
}

\end{lstlisting}
	\caption{FTI API to support transparent GPU/CPU checkpoints. FTI API
	calls and variables are marked as red.}
	\label{lst:normCkpt}
\end{listing}

Most scientific applications do not divide the same workload among the GPU and
the CPU in parallel. The computational power of a GPU device is magnitudes
larger than the one of a CPU. Assigning a workload to execute
simultaneously in both devices raises severe load balancing issues.
Hence, applications are executed in different phases, each phase is
executed by a specific device. GPUs are used for massively parallel
computational intensive phases, whereas CPUs are used for non-parallelizable,
communication heavy phases. Therefore, for a every
MPI-process, the application memory is  distributed across the main memory (CPU)
and the GPU memory. 
    \begin{figure}[t]
        \centering
        \includegraphics[width=0.5\textwidth]{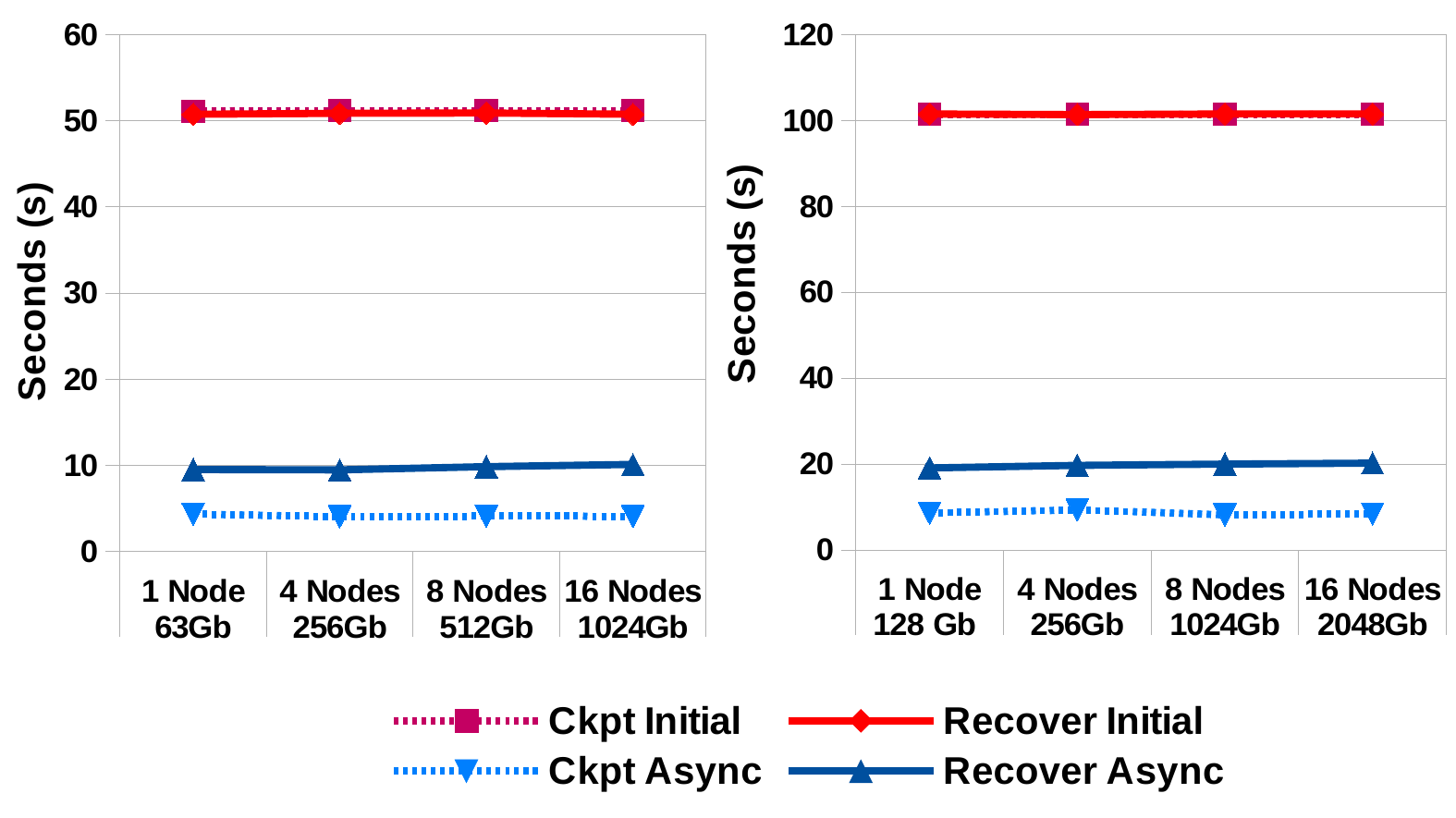}
        \caption{Execution time spent to C/R \textit{Heat2D}.}
        \label{fig:heatdis}
        \vspace{-1em}
    \end{figure}
Our target is to provide a single API to support checkpoint of different memory
regions regardless of their actual physical location. We extend a multilevel checkpoint library called FTI\cite{6114441}. FTI is a library that provides an API to the developer to efficiently perform multi-level checkpointing.  The developer uses library function calls to define which data need to be checkpointed  as well as at which execution points a checkpoint can be taken. An example using the extended GPU/CPU checkpoint API is presented in Listing \ref{lst:normCkpt}. Noticeably, in line
9 the developer allocates memory space using a unified virtual memory (UVM)
address , thus this address is accessible in the host code, whereas in line 10
the developer allocates a device pointer, which is not directly accessible
through host code. In lines 12,13,14 the developer protects three different
memory address, a host address , a UVM address and a device address however
there are no API extensions. In \textit{FTI\_Protect} the developer specifies a single address which can be either a host-memory address, a device memory address or a UVM address and the FTI runtime library will handle accordingly each different
address type. 

To support Hybrid GPU/CPU in FTI we extend the implementation of the
\textit{FTI\_Protect} API call. The function identifies the physical location of
the data.  When the checkpoint  takes place, depending on the  location of each address we perform a different action. In the case of  CPU or UVM addresses, we invoke the normal FTI C/R procedure. In the case of UVM addresses we use the CUDA driver
to fetch the data from their actual location and move them to the stable storage.
Finally, in the case of \textit{GPU} addresses, we overlap the writing of the
file with the data movement from the GPU side to the CPU side.  This is done through streams and asynchronous memory copies of chunks from GPU memory to host  memory. The reversed procedure takes place during recovery. 
After the initial implementation we performed several optimizations which achieve a speed up of $~10X$ in comparison with the initial implementation. 

We use \textit{Head2D} to test the behavior of our multi-gpu/multi-node
checkpoint methodology when the application is using UVM memory allocations. We
checkpoint \textit{Heat2D} for two different problem sizes, namely in the first
problem we checkpoint 16Gb per process whereas in the second we checkpoint 32Gb
per-process, in each node we execute 4 processes,one per GPU device, therefore the GPU devices are not shared among the processes. Finally, the problem size is weakly scaled as the number of nodes increases. When we use 16 nodes the total size of the problem size and thus the total size of the checkpointed data is equal to 1Tb and 2Tb respectively.
Figure \ref{fig:heatdis} depicts the results of our experiments for the
different methods. The \textit{x-axis} corresponds to the different problem
sizes and the different node configurations, whereas the \textit{y-axis}
corresponds to the execution time spend for the checkpoint and the recovery procedure respectively . Interestingly, the
checkpoint overhead does not increase as we increase the number of nodes for
the different problem sizes. During the checkpoint each  process
stores the data into the local NVMe , regardless of the number of nodes. The overhead decreases as we apply our optimized methods. Namely when we compare the
initial version with the async version we obtain respectively a $12.05X$ and $5.13X$ reduction in the checkpoint and recover overhead. The same amount of reduction is observed in both problem sizes, consequently our implementation strongly scales. Our initial estimations expect, for the same amount of application overhead, the extended FTI version can sustain execution in systems with 7 times smaller MTBF.

\section{Compiler: Heterogeneity- and Energy-aware Scheduler (HEATS)}
\label{sec:heats}

\begin{figure}[!t]
	\centering
	\includegraphics[scale=0.8]{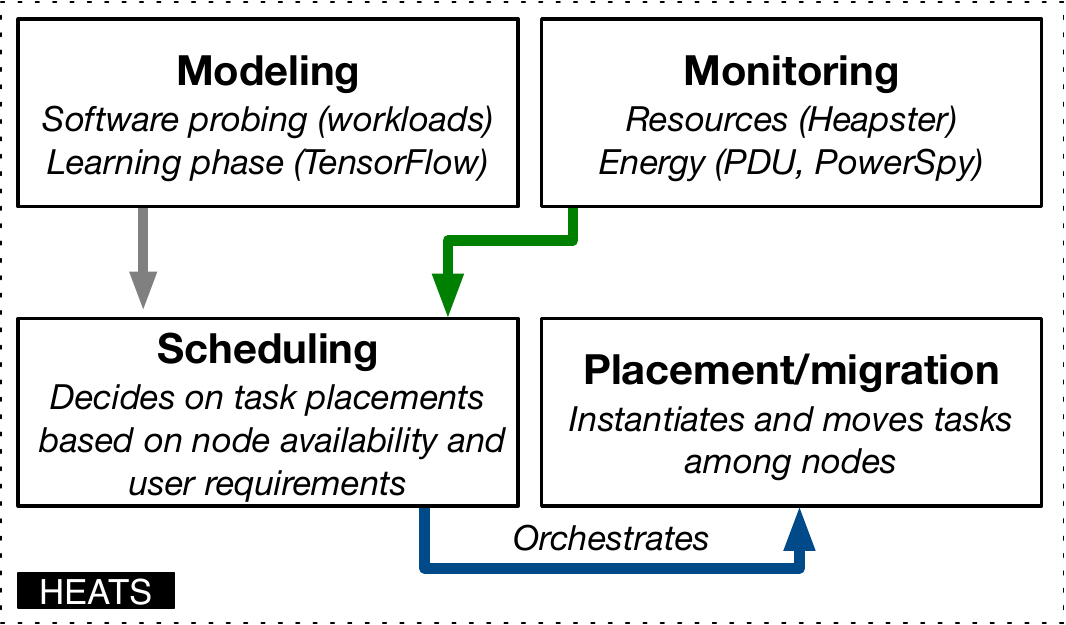}
	\caption{\sys's abstract components and interaction.}
	\label{fig:arch:components}
	\vspace{-1em}
\end{figure}

\sys is a heterogeneity- and energy-aware scheduler which allows customers to trade performance vs. energy requirements.
Our system first learns the performance and energy features of the physical hosts.
Then, it monitors the execution of tasks on the hosts and opportunistically migrates them onto different cluster nodes to match the customer-required deployment trade-offs. 

The architecture of \sys is composed of several interacting components.
Fig.~\ref{fig:arch:components} depicts these interactions. The resource requirements of a task, as for instance memory or number of cores, are specified before submission.
Resource availability in the hardware nodes is monitored and reported to \sys monitoring module.
Then, \sys computes suitable nodes for execution considering the resource requirements for all previously running tasks as well as the availability reported by the underlying system.
Next, the algorithm executes a profiling phase and estimates the performance and energy requirements of the given task in each of the previously computed available nodes.
Finally, the scheduling module relies on these estimations to compute scores for each node, to be weighted by the energy/performance ratio defined by the client.  
The best fitting node is chosen to deploy the given task.

In summary, the \sys strategy will attempt to place tasks on the most efficient host that still has enough resources to run the given task.
We define \emph{most efficient} as the closest match to the demanded energy/performance trade-off.
However, the ideal node for a task will not always be available at scheduling time. 
Therefore, we recompute our scheduling decision every now and then.
When a better fit than the current host of a task is found, the scheduler performs a migration.

The scheduling phase is triggered for the queue of all pending tasks.
The algorithm initially finds the best fit for the next task. 
It identifies its resource requirements, e.g., CPU and memory, as well as the available nodes for these resources. 
Then, it computes the score for each of the nodes.
The model is used for the profiling of nodes. 
The scores are computed by normalizing the predictions and adding the demanded weights.
Every $x$ seconds the rescheduling phase is triggered for the set of all running tasks.
If the re-execution of the best fit decides on a different target node, the task is migrated to the new host and removed from the current one~\cite{rocha2019heats}.

\section{Use Case: Smart Mirror}
\label{sec:smartmirror}

\begin{figure}
\begin{center}
\includegraphics[width=0.5\textwidth]{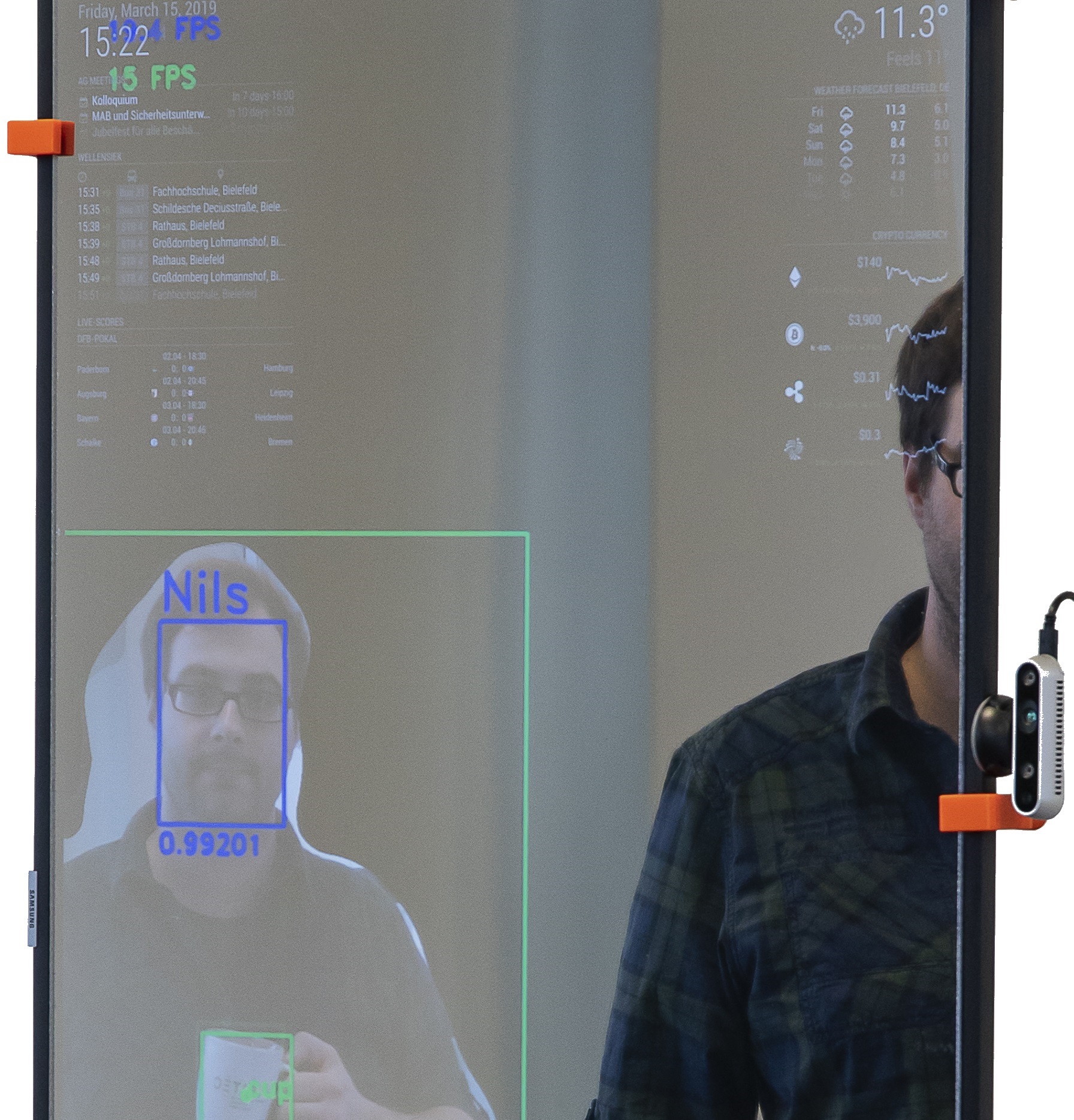}
\caption{User interface and interaction with the Smart Mirror.} 
\label{fig:smart-mirror}
\end{center}
\vspace{-1em}
\end{figure} 


An increasingly used interaction interface in smart homes is a Smart Mirror. 
Fig. \ref{fig:smart-mirror} illustrates the composition and the user interface of an example.
It consists of a semi-transparent mirror with an underlying monitor, an RGBD camera and a microphone. 
In this way, you can see your mirror image like a normal mirror and the elements shown on the display. 
In order to display personalised information such as public transport timetables, weather forecasts or menus for the respective user or to control the intelligent home environment such as temperature or lighting conditions, the most important features of a smart home are combined in this.
These are face, object, gesture, and speech recognition and are individually realized in modules underneath the MagicMirror² as the graphical overlay. Unlike other approaches, everything is thereby processed locally and no data gets into the cloud, which guarantees the personal privacy.
Neural networks like Yolov3 are providing the detections and Kalman and Hungarian filters are used to keep track.
These algorithms require a lot of computing power and have previously met on a high-end workstation with two NVIDIA GTX 1080 GPGPUs. Currently, the performance for the object, gesture, and face-detection is about 21\,FPS at 400\,W. Further optimizations on the implementation and algorithmic level including the use of specialized target architectures like FPGAs or GPU SoCs aim for a power consumption of 50\,W at 10\,FPS, which is sufficient for a seamless user experience. 

\begin{figure}
\begin{center}
\includegraphics[width=0.5\textwidth]{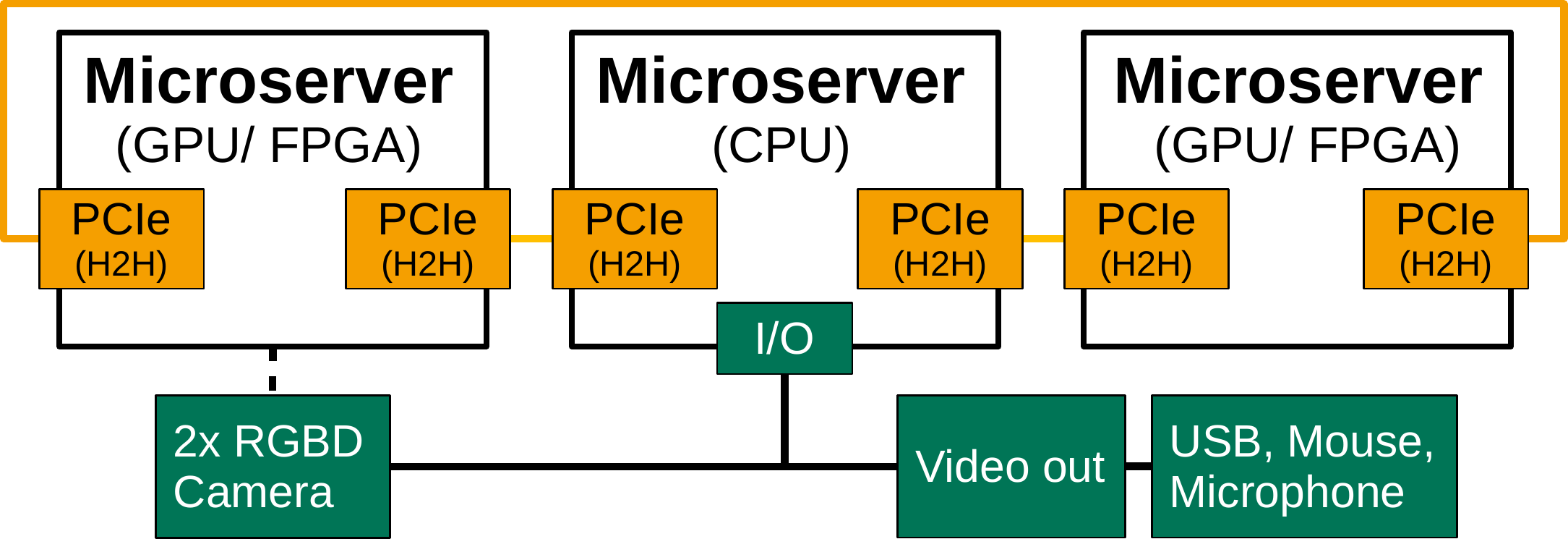}
\caption{LEGaTO Edge Server architecture optimized for Smart Mirror use-case.} 
\label{fig:edgePlatform}
\end{center}
\vspace{-1em}
\end{figure}

Based on these high demands of the Smart Mirror use-case with regard to computing power, power/energy consumption, input/outputs, integration capability into a living room, which comes with inaudible operating noise and a compact design, an optimized edge server which also supports embedded use cases is developed.
The edge server is based on 3 modular microservers utilising on the upcoming PICMG COM-HPC specification in a compact enclosure which is about 20x40~cm (Fig. \ref{fig:edgePlatform}). The modular approach allows to quickly evaluate different microserver compositions for the appliances, e.\,g., the Smart Mirror use case can be implemented with 1x~CPU + 2x~GPU or 1~CPU + 1~GPU + 1~FPGA SoC or any other microserver configuration. It should be noted that the microservers are self-sustained, the PCIe communication is used in a host-2-host fashion, such that each microserver can operate independently and is not just a PCIe peripheral of the CPU microserver. This approach supports a wide range of edge computing appliances like ADAS or Machine Learning. 
 

\section{Ongoing and Future Plans of LEGaTO} 
\label{sec:future}
Efforts in LEGaTO for the last year of the project concentrated on the optimization/developing tools/technologies aiming to achieve the final goals of the energy-saving (10X), security (10X), reliability (5X), and productivity (5X), as well as on the integration of the different components to make the full hardware-software stack tightly coupled. For instance, we are working on the extension of tools such as OmpSs for FPGA cluster, on the optimization of the backend system such as energy-aware runtime, on the integration of different LEGaTO components such as checkpointing for FPGA-based applications, and on the optimization of use cases for energy, security, reliability, and productivity.

\section{About LEGaTO}
This   project   has   received   funding   from  the   European   Union   Horizon   2020   research and   innovation   programme   under   the   LEGaTO   Project(legato- project.eu), grant agreement No 780681. The present publication  reflects  only  the  authors’  views.  The  European Commission  is  not  liable  for  any  use  that  might  be  made  of the information contained therein. The partners of the project are Barcelona Supercomputing Center (BSC, Spain), Universität Bielefeld (UNIBI, Germany), Université de Neuchâtel (UNINE, Switzerland), Chalmers Tekniska Högskola AB (CHALMERS, Sweden), Machine Intelligence Sweden AB (MIS, Sweden), Technische Universität Dresden (TUD, Germany), Christmann Informationstechnik + Medien GmbH \& Co. KG (CHR, Germany), Helmholtz-Zentrum für Infektionsforschung GmbH (HZI, Germany), TECHNION Israel Institute of Technology (TECHNION, Israel), and Maxeler Technologies Limited (MAXELER, United Kingdom).

\bibliographystyle{unsrt}
\bibliography{references}

\end{document}